 \documentclass[journal]{IEEEtran}

\ifCLASSINFOpdf
\else
   \usepackage[dvips]{graphicx}
\fi
\usepackage{url}

\usepackage{graphicx}
\usepackage{amsmath}
\usepackage{amssymb}
\usepackage{tabularray}
\usepackage{amssymb}
\usepackage{diagbox}

\usepackage{romannum}
\usepackage{verbatim}
\usepackage{multirow}
\usepackage{cite} 
\begin{document}

\title{Errorless Robust JPEG Steganography Using Steganographic Polar Codes}

\author{Jimin Zhang,
         Xianfeng Zhao,~\IEEEmembership{Senior~Member,~IEEE}, Xiaolei He 
         \thanks{  }
         \thanks{The authors are with National Engineering Laboratory for Information Security Technologies, Institute of Information Engineering, Chinese Academy of Sciences, Beijing 100085, China, and also with School of Cyber Security, University of Chinese Academy of Sciences, Beijing 100085, China (e-mail: zhangjimin@iie.ac.cn; zhaoxianfeng@iie.ac.cn; hexiaolei@iie.ac.cn).}
         }

\markboth{}
{Shell \MakeLowercase{\textit{et al.}}: Bare Demo of IEEEtran.cls for IEEE Journals}
\maketitle

\begin{abstract}
Recently, a robust steganographic algorithm that achieves errorless robustness against JPEG recompression is proposed. The method evaluates the behavior of DCT coefficients after recompression using the local JPEG encoder to select robust coefficients and sets the other coefficients as wet cost. Combining the lattice embedding scheme, the method is errorless by construction. However, the authors only concern with the success rate under theoretical embedding,  while the success rate of the implementation with practical steganographic codes is not verified. In this letter, we implement the method with two steganographic codes, i.e., steganographic polar code and syndrome-trellis code.  By analyzing the possibility of success embedding of two steganographic codes under wet paper embedding, we discover that steganographic polar code achieves success embedding with a larger number of wet coefficients compared with syndrome-trellis code, which makes steganographic polar code  more suitable under the errorless robust embedding paradigm. The experimental results show that the combination of steganographic polar code and errorless robust embedding achieves a higher success rate compared with the implementation with syndrome-trellis code under close security performance.
\end{abstract}


\IEEEpeerreviewmaketitle

\section{Introduction}

\IEEEPARstart{S}{teganography} means to modify the multimedia files like images, audio, and videos without introducing apparent traces, and the information can be extracted from modified files. Steganalysis means to detect the traces that the steganographic algorithms have introduced.  To improve the security performance of the most popular  multimedia type, i.e., JPEG images, some adaptive JPEG steganographic algorithms have been proposed. Usually the adaptive steganography consists of a cost function and a steganographic code. The cost functions in the JPEG domain include J-UNIWARD \cite{holub2013digital}, UED \cite{guo2012efficient}, UERD \cite{guo2015using}, JMIPOD \cite{cogranne2020steganography}, etc. In the past decade, most of steganographic algorithms have adopted  Syndrome-Trellis Code (STC) \cite{filler2011minimizing} as the steganographic code. Until recently, Steganographic Polar Code (SPC) \cite{li2020designing} and its variant \cite{fu2023high} have been proposed with a better performance compared with STC.

Nowadays, using social networks as channels for steganography can be more secure due to their multi-user interaction feature \cite{zhao2021}. However, social network channels can be lossy for JPEG compression is usually used to reduce bandwidth. Unfortunately, most adaptive steganographic methods assume that the transmission channel is errorless. To achieve robustness against compression, robust steganography has been proposed. In \cite{zhang2015jpeg, zhang2018dither},  robust steganographic methods utilizing techniques from image watermarking have been proposed. In \cite{tao2018towards, lu2020secure}, the authors achieve robustness by generating intermediate stego images using accurate channel modeling or an auto-encoder, where the coefficients of stego images are restored after channel compression. In \cite{zhao2018improving}, the authors propose to repeatedly compress  images to generate  robust images that barely change after compression, and the cost function is modified in \cite{zhang2021improving} to restore the robust domain formed by multiple compression. In \cite{zeng2022improving} the authors improve robustness by embedding on channel-processed cover images and avoiding modifying coefficients that could introduce compression errors. 

Most robust steganographic algorithms rely on reducing the error rate of the recompressed stego image to improve the success rate of extraction. In practice, however, the uncertainty of successful extraction increases the difficulty of achieving covert communication. Robust steganography that can ensure that the receiver can correctly extract the secret message on the sender side is very important. Recently, in \cite{butora2022errorless}, an errorless robust  embedding is constructed by using a local JPEG compressor to obtain the behavior of the compressed DCT coefficients to select robust coefficients, and using a lattice embedding method to resolve the correlation between coefficients. However, since the method sets the non-robust coefficients as wet coefficients, when users use steganographic codes for embedding, it is possible to change the wet coefficients, resulting in embedding failure. In the  work, the authors only give the success rate of the simulated optimal embedding, but do not detail the performance when embedding with steganographic codes. In this letter, we achieve errorless robust steganography using two steganographic codes, STC and SPC, and propose a practical robust steganographic method that enables errorless robust embedding with a high success rate.


The rest of the letter is organized as follows. Section  \Romannum{2}  gives the preliminaries of this letter. In Section \Romannum{3}, the performance of STC and SPC is analyzed and the proposed method is introduced in detail. Section \Romannum{4} presents the experimental results. Section \Romannum{5} is the conclusion.

\section{Preliminaries}
\subsection {The optimal embedding}
Suppose $\mathbf{y}$ and $\mathbf{x}$ are stego and cover with the length $n$ and  $\boldsymbol{\rho}$ is the distortion of the cover. Under the additive distortion scenario where the distortion of each changed pixel is independent, the overall distortion after embedding is:
\begin{equation}
D(\mathbf{x}, \mathbf{y})=\sum_{i=1}^n \rho\left(y_i\right).
\end{equation}
Set the possibility of changing $x_{i}$ to $y_{i}$ as $\pi\left(y_i\right)$, the message $\mathbf{m}$ with length $m$, and in binary embedding, the value of $y_{i} \in \mathcal{I}_{i} = \{x_{i}, \bar{x}_{i}\}, \bar{x}_{i} \neq x_{i}$. The  PLS (payload limited sender) problem can be written as :
\begin{align}
& \underset{\boldsymbol{\pi}}{\operatorname{minimize}} \; E_{\boldsymbol{\pi}}(D)=\sum_{i=1}^n \sum_{t_i \in \mathcal{I}_i} \pi\left(t_i\right) \rho\left(t_i\right) \label{equation1}\\
& \text { subject to } H(\boldsymbol{\pi})=-\sum_{i=1}^n \sum_{t_i \in \mathcal{I}_i} \pi\left(t_i\right) \log _2 \pi\left(t_i\right)=m. \label{equation2}
\end{align}
Solving the optimization problem by the maximum entropy principle, the optimal distribution $\boldsymbol{\pi}_{\lambda}$ is 
\begin{equation}
\pi_\lambda\left(y_i\right)=\frac{\exp \left(-\lambda \rho\left(y_i\right)\right)}{\sum_{t_i \in \mathcal{I}_i} \exp \left(-\lambda \rho\left(t_i\right)\right)}, \quad 1 \leq i \leq n.
\end{equation}

\subsection {Errorless robust steganography}
To ensure errorless robust embedding, robust steganography should preserve the coefficients of a stego image after compression. To achieve this goal, the method in \cite{butora2022errorless} generates a robust set $\mathcal{R}$ with the output of a JPEG encoder. A coefficient in the robust set satisfies three requirements: it does not change processed coefficients, it preserves the modification, and it preserves its value if there is no change. If a coefficient does not satisfy these requirements, it belongs to the non-robust set with the cost of modification as infinity, and the value after recompression is used as the embedding domain. Considering the influence of different modifications in the same DCT block, the method divides an image into 64 lattices by the mode in a DCT block. Because in a lattice the coefficients are from different blocks, the cost of the lattice can be acquired by only three times compression. After calculating costs for all coefficients, $\lambda$ can be calculated using (\ref{equation2}). The payload in each lattice can be calculated with $\lambda$ and the costs of coefficients in the lattice. The cost calculation and embedding are then performed sequentially in each lattice.

\subsection {STC and SPC embedding}
\begin{figure}[htbp]
      \centering
      \includegraphics[width=0.8\linewidth]{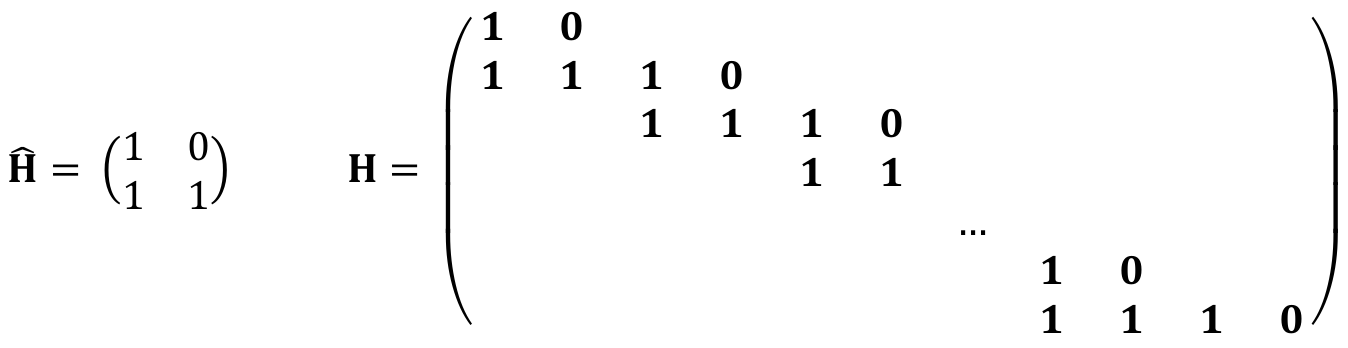}
      \caption{The structure of the STC parity-check matrix.}
      \label{figure:stc}
\end{figure}

STC uses the standard trellis representation of convolutional codes to embed messages. In STC, the parity-check matrix $\mathbf{H} \in\{0,1\}^{m \times n}$ of length $n$ and codimension $m$ is obtained by placing a small submatrix $\hat{\mathbf{H}}$ of size $h \times w$ as in Fig. \ref{figure:stc},  
where $h$ is called constraint heights and $w$ is equal to $\lfloor n/m \rfloor$. Each $\mathbf{z} \in\{0,1\}^n$  satisfying  $\mathbf{H}\mathbf{z}=\mathbf{m}$ is represented as a path in the syndrome trellis. In the forward Viterbi algorithm, the path with the smaller cost is chosen, and in the backward Viterbi algorithm, the stego image is obtained.

A polar code is specified by $\left(n, k, \mathcal{A}, \mathbf{u}_{\mathcal{A}^c}\right)$. Set $\mathcal{A}$ of dimension $k (k < n)$ is the set of information indices  carrying information $\mathbf{u}_{\mathcal{A}}$, and indices set $\mathcal{A}^c$ carries frozen bits  $\mathbf{u}_{\mathcal{A}^c}$ of dimension $ n - k$. The choice of $\mathcal{A}^{c}$ corresponds to the section of $ n - k$ worst polarized channels. The source code $\mathbf{u}$ is formed by $\mathbf{u} = (\mathbf{u}_{\mathcal{A}}, \mathbf{u}_{\mathcal{A}^{c}})$, and the codeword $\mathbf{c}$ is generated by $\mathbf{c} = \mathbf{u}\mathbf{G}_{n}$. The generator matrix is $\mathbf{G}_n=\mathbf{B}_n \mathbf{F}^{\otimes s}$ for $n = 2^{s}$, where $\mathbf{B}_{n}$ is the bit-reversal permutation matrix, $\mathbf{F}^{\otimes s}$ denotes the $k_{th}$  Kronecker power of $\mathbf{F}$, and $\mathbf{F} \triangleq \left[\begin{array}{l}
1,0 \\
1,1
\end{array}\right]$. The generator matrix of $s=2$ is 
\begin{equation}
\label{GSPC}
\mathbf{G}_{4} = \left[ \begin{array}{l}
1, 0, 0, 0 \\
1, 0, 1, 0\\
1, 1, 0, 0\\
1, 1, 1, 1
\end{array}
\right].
\end{equation}

The embedding process of SPC in \cite{li2020designing} is firstly calculating the Bhattacharyya parameters $\mathbf{Z}$ as:
\begin{equation}
   \left\{\begin{array}{l}
Z\left(W_n^{(2 j-1)}\right)=2 Z\left(W_{n / 2}^{(j)}\right)-Z\left(W_{n / 2}^{(j)}\right)^2 \\
Z\left(W_n^{(2 j)}\right)=Z\left(W_{n / 2}^{(j)}\right)^2, \quad 1 \leq j \leq n / 2
\end{array}\right. 
\end{equation}
where $Z\left(W_n^{(1)}\right) = \alpha$ and $\alpha = m/n$ is the payload.
The positions of $m$ channels with the largest $Z$ values are chosen as the frozen indices $\mathcal{A}^{c}$.  Then $\lambda$ is solved by (3), and the initial Log-Likelihood Ratio (LLR) is calculated as
\begin{equation}
    L_1^{(1)}\left(x_i\right)=\left(2 \mathcal{P}\left(x_i\right)-1\right) \cdot \ln \frac{\pi_\lambda\left(\bar{x}_i\right)}{1-\pi_\lambda\left(\bar{x}_i\right)}, \quad 1 \leq i \leq n
\end{equation}
where $\mathcal{P}(x)$ returns the parity of $x$. Then $\mathbf{u}_{\mathcal{A}^{c}}$,  $\mathbf{L}_1^{(1)}$, list size $l$, $\mathcal{A}^{c}$, and $\mathcal{P}(\mathbf{x})$ are input into the list version of the successive cancellation decoder (SCL), where $\mathbf{u}_{\mathcal{A}^{c}} = \mathbf{m}$, which generates $\mathbf{u}$. Then the stego image $\mathbf{y}$ can be obtained by $\mathbf{u}\mathbf{G}_n$, and the process is denoted as $\mathbf{y} = $ SCL$(\mathbf{u}_{\mathcal{A}^{c}},\mathcal{A}^{c},\mathbf{L}_1^{(1)},l, \mathcal{P}(\mathbf{x}))$. The information can be extracted by $\mathcal{P}(\mathbf{y})\mathbf{G}_n$ and $\mathcal{A}^{c}$. The list size $l=6$ in this letter. In \cite{fu2023high}, the steganographic polar code with high computational efficiency is developed based on sub-polarized channels.

\section{Proposed method}
\subsection{The analysis of SPC and STC on wet paper embedding}
In errorless robust steganography, the success rate depends on the ability of embedding codes to avoid embedding wet coefficients.  When polar codes are used for steganographic embedding,  in order to have the highest probability of finding a solution in wet paper embedding,  the frozen bits used to embed  messages should
correspond to  columns with most 1s in the generation matrix.  Under this condition, in \cite{diouf2017performances} the authors prove that when the message length is $log_{2}n + 1$, it can always find solutions in  wet paper embedding with polar codes if the number of wet points is less or equal to $n/2 -1$. In STC embedding, at most $h \times w $ cover elements can be changed in order to embed one bit of information. Assuming in STC or polar code-based steganographic codes, the minimum of the length of cover elements used to embed one bit of messages is $l_{m}, l_{m} \geq 1$, and the number of wet coefficients is $w_n$, then there are solutions without changing wet coefficients if $w_n \leq l_{m} - 1$. The proof is as follows, where the values of $\mathbf{x}$, $\mathbf{y}$, and $\mathbf{m}$ are in the binary domain:
\begin{itemize}
\item [1)] When $l_{m}=1$ and $w_n=0$, there is no wet coefficient to be changed when embedding.

\item [2)] Assume that when $l_{m} = k$ and $w_n \leq k - 1$, a solution can be found without changing wet coefficients.  When $l_m = k + 1$ and $w_n \leq k $, suppose cover element $x_{i}$ is the wet coefficient and it is used to embed message $m_{j}$ with other $k$ cover elements. Then delete $x_{i}$  and replace  $m_{j}$  with $m_{j} \oplus x_{i}$. If there are messages bits $\mathcal{B}$ that are also embedded using $x_{i}$, replace them with $m_p \oplus x_{i}, m_p \in \mathcal{B}$ and remove $x_{i}$. The operations make $l_{m} = k$ and  $w_n \leq k - 1$. Since we assume that a solution can be found in this case,  when $l_{m} = k + 1$ and  $w_{n} \leq k$, the solution can be found without changing wet coefficients.
\end{itemize}

\begin{figure}[htbp]
      \centering
      \includegraphics[width=0.6\linewidth]{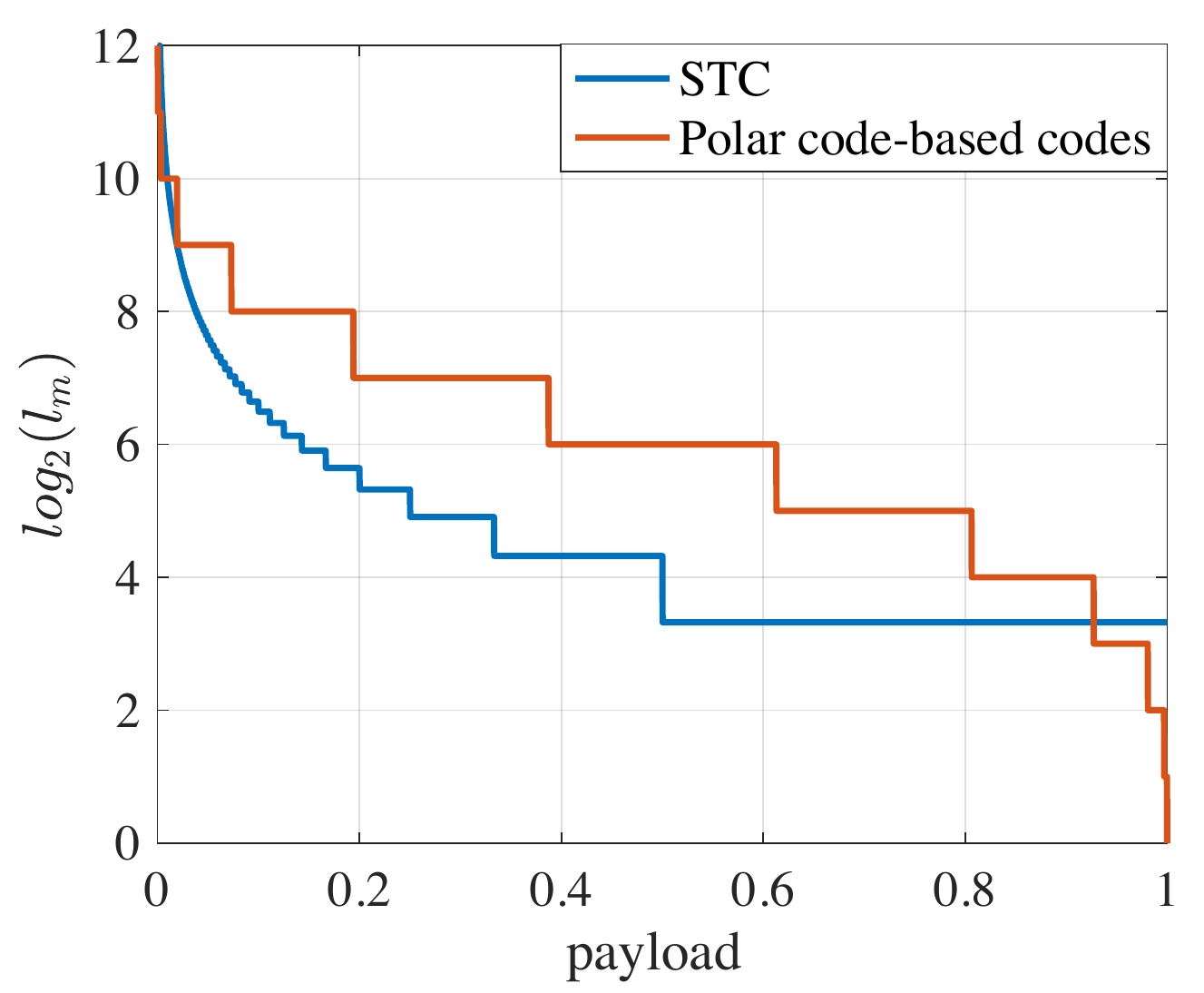}
      \caption{The $log_{2}(l_{m})$ when embedding with different codes and  payloads $(m/n)$ with $n = 2^{12}$.}
      \label{figure:MCL}
\end{figure}

When embedding the same length of messages, the coding scheme with larger $l_{m}$ can find a solution in wet paper embedding  with the tolerance of larger $w_n$. when $n=2^{12}$, assuming steganographic codes based on polar codes  choose the best positions for embedding and the $l_{m}$ in STC is min$(\lfloor n/m \rfloor \times h,n)$ and $h=10$, the values of $l_{m}$ of two codes with  different payloads ($m/n$) are shown  in Fig. \ref{figure:MCL}. The result presents that $l_{m}$ of  polar code-based codes is larger when the  payload is in the range of [0.0092, 0.927], which demonstrates that when the payload is not too high or too low,  polar code-based codes can avoid wet coefficients changing with a larger $w_n$. In the errorless robust embedding scheme, polar code-based  codes have a higher success rate because the solution can be found with more non-robust coefficients compared with the implementation with STC.

\begin{figure}[htbp]
      \centering
      \includegraphics[width=0.96\linewidth]{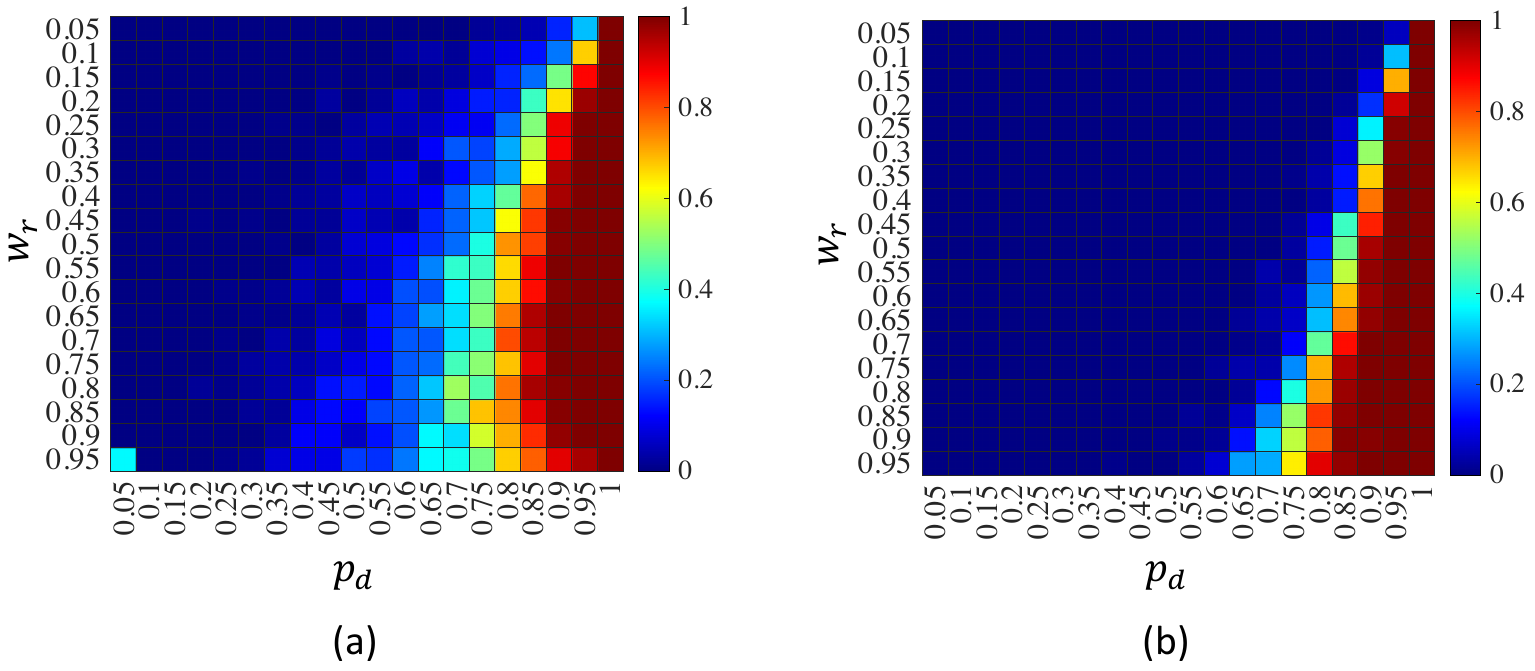}
      \caption{
The value of $P_{w}$ when embedding with different $w_r$ and $p_d$ under the constant cost profile. (a) embedding with STC, (b) embedding with SPC.
}
      \label{figure:hotmapallone}
   \end{figure}

\begin{figure}[htbp]
      \centering
      \includegraphics[width=0.96\linewidth]{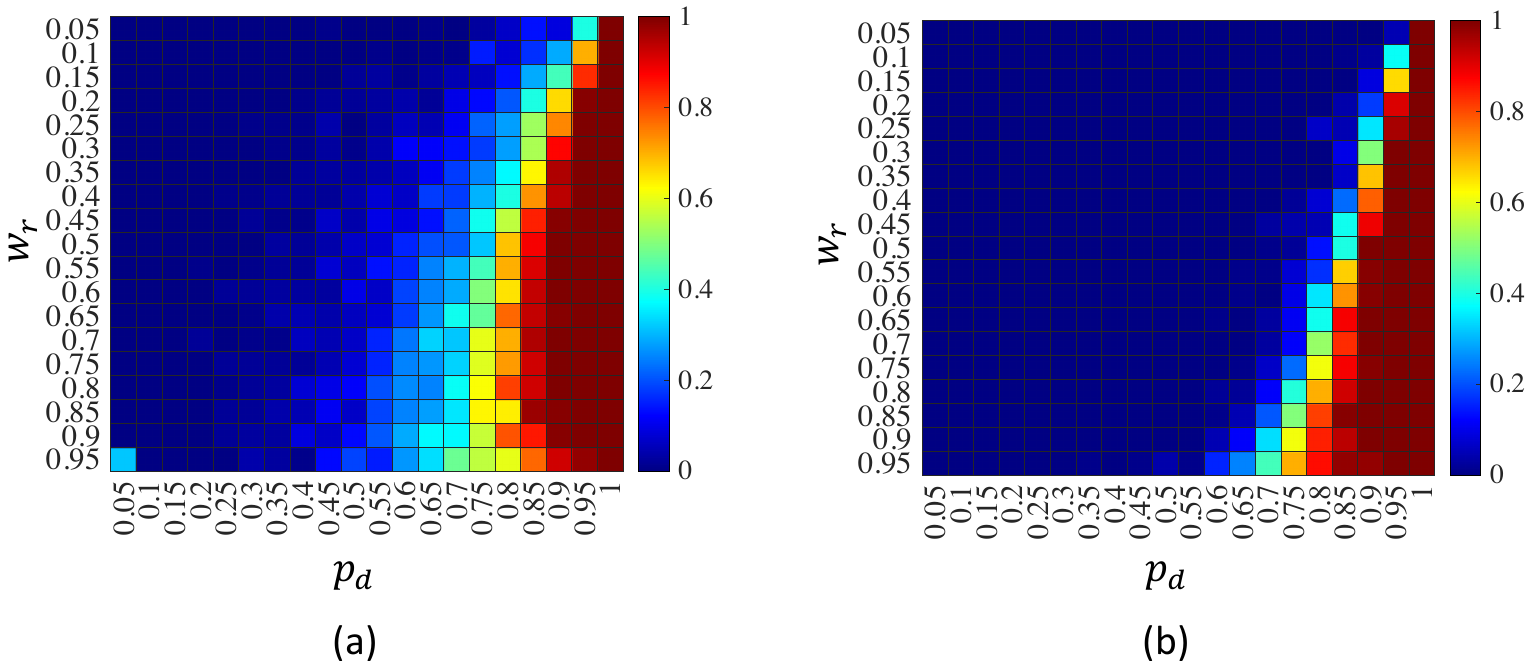}
      \caption{The value of $P_{w}$ when embedding with different $w_r$ and $p_d$ under the square cost profile. (a) embedding with STC, (b) embedding with SPC.}
      \label{figure:hotmapsquare}
   \end{figure}

Suppose the wet ratio $w_r=w_n/n$, and the payload on dry coefficients $p_d = m/(n-w_n)$. In order to obtain the possibility of changing wet coefficients $P_{w}$ in the actual embedding, the values of $P_{w}$ under different $w_r$ and $p_d$ are tested using SPC and STC with $n = 2^{12}$. In each case the random messages are embedded by binary embedding with random cover and random wet coefficients positions 100 times. Because ternary embedding can be split into two binary embedding, the results are easily extended to ternary embedding. The results under the constant cost profile and the square cost profile are shown in Fig. \ref{figure:hotmapallone} and Fig. \ref{figure:hotmapsquare}, respectively. It can be seen  the results under different cost profiles are similar. There are more cases where $P_{w}=0$ in SPC embedding compared with STC, and most of them are located in cases where $p_d$ is close to its median value. When $p_d$ is close to 1, neither can avoid changing wet coefficients, because  $l_{m}$ decreases as $p_d$ increases. When the value of $w_r$ and $p_d$ are both high, there are  cases where $P_{w}$ of STC is slightly lower than SPC, but the effect is small because $P_{w}$ of both codes is large ($P_{w} \geq 0.49$).

\begin{figure*}[htbp]
      \centering
      \includegraphics[width=0.90\linewidth]{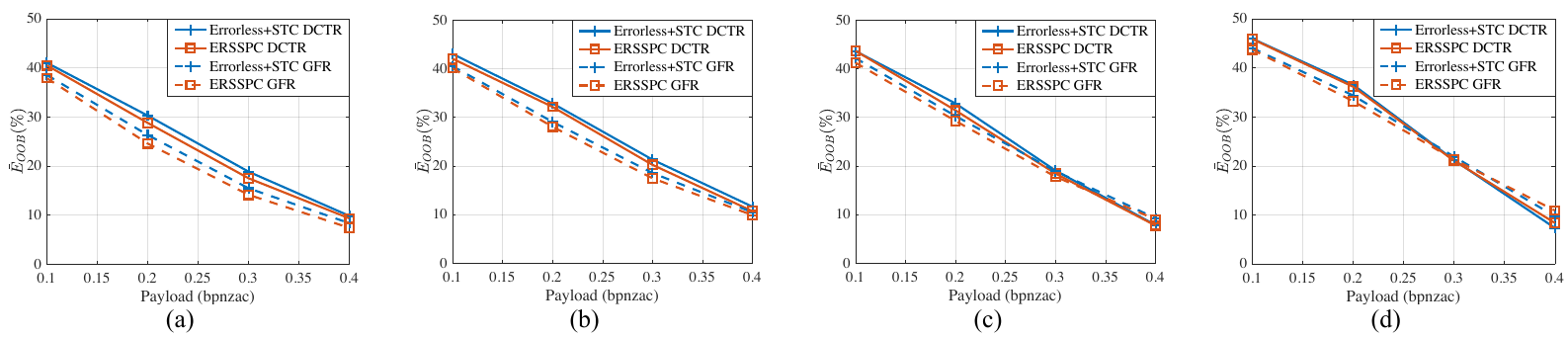}
      \caption{Security performance of proposed ERSSPC and errorless robust embedding with STC. (a) QF = 75, (b) QF = 85, (c) QF = 90, (d) QF = 95.}
      \label{figure:security}
   \end{figure*}

\begin{figure}[htbp]
      \centering
      \includegraphics[width=0.90\linewidth]{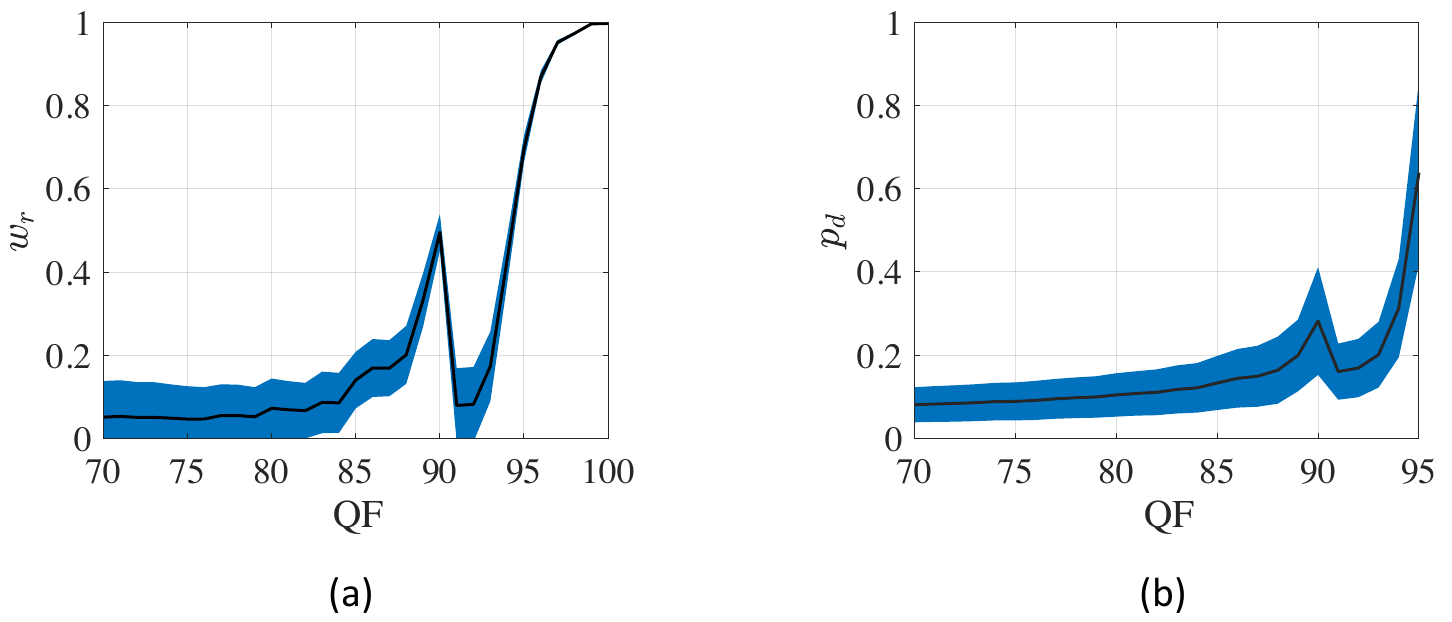}
      \caption{
The mean (black line) and standard deviation (blue area) of $w_r$ and $p_{d}$ when embedding the last lattice with the random scanning strategy under different quality factors and the payload 0.5 bpnzac. The compressor is ImageMagick's \texttt{convert}. (a) mean$(w_r)$ and std$(w_r)$, (b) mean$(p_d)$ and std$(p_d)$.}
      \label{figure:qfwetpayload}
   \end{figure}

When errorless robust steganography is performed, the mean and standard deviation value of $w_r$ and $p_d$ in the last lattice when embedding with the random scanning strategy and the payload is set as 0.5 bpnzac are shown in Fig. \ref{figure:qfwetpayload}. Images are compressed with \texttt{convert} in ImageMagick under different quality factors (QFs). When QF $>95$, $p_d$ is not drawn for its mean  is greater than 1. The results present that $w_r$ and $p_d$ generally increase as the quality factor increases. when QF $=95$, mean$(w_r) = 0.7$ and mean$(p_d) = 0.63$ with std$(p_d) = 0.22$, there is the possibility that STC performs better than SPC. From Fig. \ref{figure:hotmapsquare}, it can be observed that in other cases, the value of $P_{w}$ of SPC is mostly lower than STC.

\subsection{The embedding process}
Since  steganographic polar code has a higher possibility of success in wet paper embedding, we propose a method called Errorless Robust Steganography with SPC (ERSSPC). The embedding process is similar with \cite{butora2022errorless} and described as follows. First, the initial robust set $\mathcal{R}^{ini}$ is calculated, and the message $\mathbf{m}$ is split with initial costs $\mathbf{\rho}^{ini,k}$ in each lattice. The inital pseudo-stego $\mathbf{y}^{s,0} = \mathbf{x}$. Then the lattices are embedded sequentially. When we embed the message $\mathbf{m}^{k}$ in the $k$th lattice $\mathbf{x}^{k}$, the robust set $\mathcal{R}^{k}$ is calculated based on the pseudo-stego $\mathbf{y}^{s,k-1}$. The adaptive costs are updated as $\rho^{k}_{i} = + \infty$, and $x_{i}^{k} = $ C$(\mathbf{y}^{s,k-1})_{i}^{k}$, if $x_{i}^{k} \notin \mathcal{R}^{k}$ and C$(\cdot)$ means the JPEG compression function. The cover and costs are randomly permutated to $\mathbf{x}^{k\prime}$ and $\mathbf{\rho}^{k\prime}$. Then $\mathcal{A}^{c}$ and $\mathbf{L}^{(1)}_{1}$ are calculated with (6) and (7). After that $\mathbf{y}^{k\prime}$ is generated by SCL$(\mathbf{u}_{\mathcal{A}^{c}}, \mathcal{A}^{c}, \mathbf{L}^{\left(1\right)}_{1}, l, \mathcal{P}(\mathbf{x}^{k\prime}))$, where $\mathbf{u}_{\mathcal{A}^{c}} = \mathbf{m}^{k}$, and $\mathbf{y}^{k}$ is generated by inverse permutation. Then $\mathbf{y}^{k}_{i} = (\mathbf{y}^{s,k-1})_{i}^{k}$, if $x_{i}^{k} \notin \mathcal{R}^{k}$, and the pseudo-stego $\mathbf{y}^{s,k-1}$ is updated to $\mathbf{y}^{s,k}$ with $\mathbf{y}^{k}$. After 64 lattices are embedded and wet coefficients are not modified,  the embedding successes and the stego image $\mathbf{y}=\mathbf{y}^{s,64}$ is obtained.

\section{Experimental result}

\subsection{Experimental setup}
In this part, the testing images are $2,000$ grayscale images with  size  $512 \times 512$ randomly selected from BossBase ver 1.01 \cite{bas2011break}. The cover images are generated by JPEG compression using ImageMagick's \texttt{convert} with quality factors 75, 85, 90, and 95. The steganalysis algorithms used for testing the security performance are DCTR \cite{holub2014low} and GFR \cite{song2015steganalysis} features equipped with ensemble classifiers. The performance is evaluated with ten tests with random splits of testing images and the mean  value of error rate  $E_{oob}$ in every test that minimizes overall detection error rate  is used as the quantification of security performance. The cost function used is UERD \cite{guo2015using} for its computational efficiency and satisfying security performance.

\subsection{Success rate and security performance}

The successful embedding means that the image can successfully embed  secret messages and the information can be correctly extracted. The robust embedding used for comparison is errorless robust embedding implemented with STC (Errorless+STC) and $h=10$.  The results in Table \ref{table_success} show that the STC always has the possibility of embedding failure. The proposed ERSSPC can successfully embed messages without embedding on  wet coefficients when  QF $ = 75$, $85$, and $90$. Because of the high value of $w_{r}$ and $P_{d}$ when QF $ = 95$, the proposed method has a small chance of embedding failure, but its success rate is still higher than STC. When the payload is 0.4 bpnzac and QF $=95$, the large gap between the two methods  is caused by the fact that  $p_d$ at this time is distributed around the median value of $p_d$ in Fig. \ref{figure:hotmapsquare}.

\begin{table}
\centering
\caption{ Success rate (\%) of proposed ERSSPC and errorless robust  embedding with STC with different quality factors and payloads (bpnzac).}
\label{table_success}
\resizebox{0.8\linewidth}{!}{
\begin{tabular}{|c|c|c|c|c|c|} 
\hline
\multirow{2}{*}{Method} & \multirow{2}{*}{QF} & \multicolumn{4}{c|}{Payload (bpnzac)}                            \\ 
\cline{3-6}
                        &                     & 0.1            & 0.2          & 0.3            & 0.4             \\ 
\hline
\multirow{4}{*}{\begin{tabular}[c]{@{}c@{}}Errorless\\ +STC\end{tabular}} & 75                  & 97.95          & 99.80        & 99.85          & 99.90           \\ 
\cline{2-6}
                        & 85                  & 99.50          & 99.90        & 99.95          & 99.95           \\ 
\cline{2-6}
                        & 90                  & 99.80          & 99.80        & 99.75          & 98.50           \\ 
\cline{2-6}
                        & 95                  & 99.90          & 99.10        & 95.15          & 91.95           \\ 
\hline
\multirow{4}{*}{ERSSPC} & 75                  & \textbf{100}   & \textbf{100} & \textbf{100}   & \textbf{100}    \\ 
\cline{2-6}
                        & 85                  & \textbf{100}   & \textbf{100} & \textbf{100}   & \textbf{100}    \\ 
\cline{2-6}
                        & 90                  & \textbf{100}   & \textbf{100} & \textbf{100}   & \textbf{100}    \\ 
\cline{2-6}
                        & 95                  & \textbf{99.95} & \textbf{100} & \textbf{99.95} & \textbf{99.80}  \\
\hline
\end{tabular}}
\end{table}

The security performance of proposed ERSSPC and Errorless+STC is shown in Fig. \ref{figure:security}. From results we can discern that the security performance of ERSSPC is close to Errorless+STC. The maximum drop of the proposed method is $1.74\%$ compared with Errorless+STC when QF $= 75$ and the payload is $0.2$ bpnzac detected by GFR features. The results verify that the proposed method improves the success rate with a small decrease in  security performance.

\section{CONCLUSION}
In this letter we proposed a robust JPEG steganographic method based on the errorless robust embedding scheme and SPC. We analyzed the polar code-based steganographic codes can more easily avoid changing wet coefficients and verified it in the actual embedding scenario. In the experiments we showed the success embedding rate of the proposed method is higher compared with the implementation of STC with close security performance. Further research will focus on the application of the proposed method on more complex compression channels and  efficiency improvement.
\bibliographystyle{IEEEtran}
\bibliography{IEEEabrv,ref}
\end{document}